\documentclass[aps,prl,preprint,groupedaddress,nofootinbib,floatfix]{revtex4-1}
\usepackage{graphicx,multirow,xspace,amsmath,amssymb}

\newcommand{\sqsn}{\mbox{$\sqrt{s_{\mathrm{NN}}}$}\xspace}

\def\mean#1{\ensuremath{\left<#1\right>}}

\newcommand{\UA}{\ensuremath{U_{A}(1)}\xspace}
\newcommand{\lambdas}{\ensuremath{\lambda_{*}}\xspace}
\newcommand{\lambdasmax}{\ensuremath{\lambda_{*}^{\mathrm{max}}}\xspace}
\newcommand{\lamfrac}{\ensuremath{\lambdas(\mT)/\lambdasmax}\xspace}
\newcommand{\mT}{\ensuremath{m_\mathrm{T}}\xspace}
\newcommand{\pT}{\ensuremath{p_\mathrm{T}}\xspace}
\newcommand{\etap}{\ensuremath{{\eta^\prime}}\xspace}

\newcommand{\metap}{\ensuremath{m_{\etap}}\xspace}
\newcommand{\meps}{\ensuremath{m_{\etap}^{*}}\xspace}
\newcommand{\Tfo}{\ensuremath{T_{\mathrm{FO}}}\xspace}
\newcommand{\Teff}{\ensuremath{T_{\mathrm{eff}}}\xspace}
\newcommand{\Tcond}{\ensuremath{T_{\mathrm{cond}}}\xspace}
\newcommand{\Binv}{\ensuremath{B^{-1}}\xspace}
\newcommand{\uT}{\ensuremath{\mean{u_\mathrm{T}}}\xspace}
\newcommand{\GeV}{\ensuremath{\mathrm{GeV}}\xspace}
\newcommand{\MeV}{\ensuremath{\mathrm{MeV}}\xspace}
\newcommand{\pip}{\ensuremath{\pi^+}\xspace}
\newcommand{\pim}{\ensuremath{\pi^-}\xspace}

\begin{document} 


\title{Indirect observation of an in-medium $\etap$ mass reduction \\ in $\sqrt{s_{NN}}=200$ GeV Au+Au collisions}

\author{T.~Cs\"org\H{o}}
\affiliation{Dept. Physics, Harvard University, 17 Oxford St, Cambridge, MA 02138, USA}
\affiliation{MTA KFKI RMKI, H--1515 Budapest 114, P.O.~Box~49, Hungary}

\author{R. V\'ertesi}
\affiliation{MTA KFKI RMKI, H--1515 Budapest 114, P.O.~Box~49, Hungary}

\author{J.~Sziklai}
\affiliation{MTA KFKI RMKI, H--1515 Budapest 114, P.O.~Box~49, Hungary}

\date{\today}
\vfill
\begin{abstract} 
PHENIX and STAR data on the intercept parameter of 
the two-pion Bose-Einstein correlation functions in $\sqsn=200\ \GeV$ Au+Au collisions
are analyzed in terms of different models of hadronic multiplicities.
To describe this combined PHENIX and STAR  dataset, 
an in-medium $\eta^\prime$ mass reduction of at least 200 MeV is needed, at the 99.9 \% confidence
level in the considered model class. 
Such a significant \etap mass modification may indicate the restoration of the \UA symmetry in a hot and dense hadronic matter and the return of the 9th ``prodigal'' Goldstone boson.
\end{abstract}
\vfill\null
\maketitle

Although the quark model exhibits a $U(3)$ chiral symmetry in the limit of massless up, down and strange quarks,
and in principle 9 massless Goldstone modes are expected to appear when this symmetry is broken, only 8 light pseudoscalar mesons are observed experimentally.
This puzzling mystery is resolved by the Adler-Bell-Jackiw \UA anomaly: instantons tunneling between topologically different QCD vacuum states explicitely break the \UA part of the $U(3)$ symmetry. Thus the 9th Goldstone boson is expected to be massive, and is associated with the \etap meson, which has a mass of 958 MeV, approximately twice that of the other pseudoscalar mesons.
In high energy heavy ion collisions, where a hot and dense medium is created, the \UA
symmetry of the strong interactions  may temporarily be restored \cite{kunihiro,kapusta,huang},
even below the critical temperature for quark deconfinement~\cite{Fodor:2009ax}.
Thus the mass of the $\etap(958)$ mesons may be reduced to its  quark model value of about 500 MeV, 
corresponding to the return of the ``prodigal" 9th Goldstone boson~\cite{kapusta}. 
Here we report on an indirect observation of such an in-medium $\etap$ mass modification based 
on a detailed analysis of PHENIX and STAR charged pion Bose-Einstein correlation (BEC) data~\cite{phnxpub,starpub}.

The abundance of the $\etap$ mesons with reduced mass may be increased at low \pT, 
by more than a factor of 10. One should emphasize that the \etap (and $\eta$) mesons almost 
always decay after the surrounding hadronic matter has frozen out, due to their small 
annihilation and scattering cross sections, and their decay times that are much longer than the
characteristic 5-10 fm/c decoupling times of the fireball created in high energy heavy ion 
collisions. Therefore one cannot expect a direct observation of the mass shift of the $\etap$ 
(or $\eta$) mesons: all detection possibilities of their in-medium  mass  modification have 
to rely on their enhanced production.
An enhancement of low transverse momentum \etap mesons contributes to an enhanced production of soft charged pions mainly through the $\etap \rightarrow \eta + \pi^+ + \pi^- \rightarrow (\pi^+ + \pi^0 + \pim) + \pip + \pim$ decay chain and also through other, less prominent channels. As the \etap decays far away from the fireball, the enhanced production of pions in the corresponding halo region will reduce the strength of the Bose-Einstein correlation between soft charged pions.
The transverse mass ($\mT = \sqrt{m^2 + \pT^2}$) dependence of the extrapolated intercept parameter 
\lambdas of the charged pion Bose-Einstein correlations 
was shown to be an observable that is sensitive to such an enhanced $\etap$ multiplicity \cite{vance}. 
The proposed decrease of $\lambdas(\mT)$ data at low transverse mass 
has been observed both by PHENIX~\cite{phnxpub} and STAR~\cite{starpub,Abelev:2009tp}.

Our main analysis tool was a Monte-Carlo simulation of the transverse mass dependence of
the long lived resonance multiplicities including the possibility of an enhanced $\etap$
production at low transverse momentum, due to a partial in-medium \UA restoration and a related $\etap$ mass modification. 
This model and the related reduction of the effective intercept parameter of the two-pion Bose-Einstein
correlation function was proposed first in ref.~\cite{vance} and detailed recently in refs.~\cite{Vertesi:2009ca,Vertesi:2009wf}.

In thermal models, the production cross sections of the light mesons are exponentially suppressed by the mass. 
Hence one expects about two orders of magnitude less \etap mesons from the freeze-out than pions. This suppression, however, may be moderated as a consequence of a possible \etap mass reduction, and the $\etap$ mesons may show up in an enhanced number. 
The number of in-medium $\etap$ mesons is calculated with an improved Hagedorn formula
yielding the following \etap enhancement factor:
\begin{equation}\label{eq:prietamtdist}
f_\etap=\left(\frac{\meps}{\metap}\right)^\alpha e^{- \frac{\metap-\meps}{\Tcond}}.
\end{equation}
This formula includes a prefactor with an expansion dynamics dependent exponent $\alpha\approx 1-d/2$ for an expansion in $d$ effective dimensions~\cite{Csorgo:1995bi}. As a default value, $\alpha = 0$ was taken~\cite{vance} and, for the systematic investigations,
this parameter was varied between $ -0.5 \le \alpha \le 0.5$.
Other model parameters and their investigated ranges  are described as follows:
$\Tcond$ in the above formula corresponds to 
the temperature of the medium when the in-medium modified $\etap$ mesons are formed; its default value
was taken to be $\Tcond = 177$ MeV~\cite{vance} and varied systematically
between 140 and 220 MeV. 
Resonances with different masses were simulated with a mass dependent slope parameter
$\Teff = \Tfo + m \langle u_T\rangle^2 $, where the default values of $\Tfo = 177$ MeV
and $\langle u_T\rangle = 0.48$~\cite{Adler:2003cb} were utilized and 
systematically varied in the range of 100 MeV  $\le \Tfo \le $ 177 MeV and 0.40  $\le \langle u_T\rangle \le $ 0.60~.

Once produced, the \etap is expected to be decoupled from other hadronic matter, since its annihilation and scattering cross sections are very small~\cite{kapusta}.
%
%
If the \etap mass is reduced in the medium, the observed \etap spectrum will consist of two components.
If the \pT of the \etap is large enough, it can get on-shell and escape. This will produce a thermal component of the spectrum. 
Energy conservation at mid-rapidity implies ${m_{\etap}^*}^2+{p_{T,\etap}^{*}}^2={m_{\etap}}^2+{p_{T,\etap}}^2$.
(In the latter equation the quantities marked with an asterisk denote the properties of the in-medium \etap, while the ones without an asterisk refer to the free \etap.)
On the other hand, \etap-s with ${p_{T,\etap}^{*}} \le \sqrt{ {m_{\etap}^*}^2-{m_{\etap}}^2 }$ will not be able to leave the hot and dense region through thermal fluctuation since they cannot compensate for the missing mass~\cite{kapusta,huang}, and thus will be trapped in the hot and dense region until it disappears. As the energy density of the medium is dissolved, the effect of QCD instantons increases and the trapped \etap mesons regain their free mass and appear at low \pT. 

Previous simulations~\cite{vance} considered the trapped \etap mesons to leave the dissolving medium with a negligible \pT, which resulted in a steep hole in the extrapolated intercept parameter $\lambdas(\mT)$ at a characteristic transverse mass of $\mT\le 250\ \MeV$~\cite{vance,phnxpre}. In this simplified scenario the only free parameter was the in-medium \etap mass, determining the depth of the observed hole. 
In the present analysis the \etap-s from the decaying condensate are given a random transverse momentum, following Maxwell-Boltzmann statistics with a newly introduced inverse slope parameter \Binv{}, which is necessary to obtain a quality description of the width and the slope of the $\lambdas(\mT)$ data of PHENIX and STAR in the $\mT \approx 300$ MeV region.
Physically, $\Binv$ is limited by $\Tfo$, so the trapped \etap -s may gain only moderate transverse momenta.
Hence the enhancement will mostly appear at low $p_T$~\cite{kunihiro,kapusta,huang}.

\begin{figure}[tp]
\begin{center}
\includegraphics[width=.9\columnwidth]{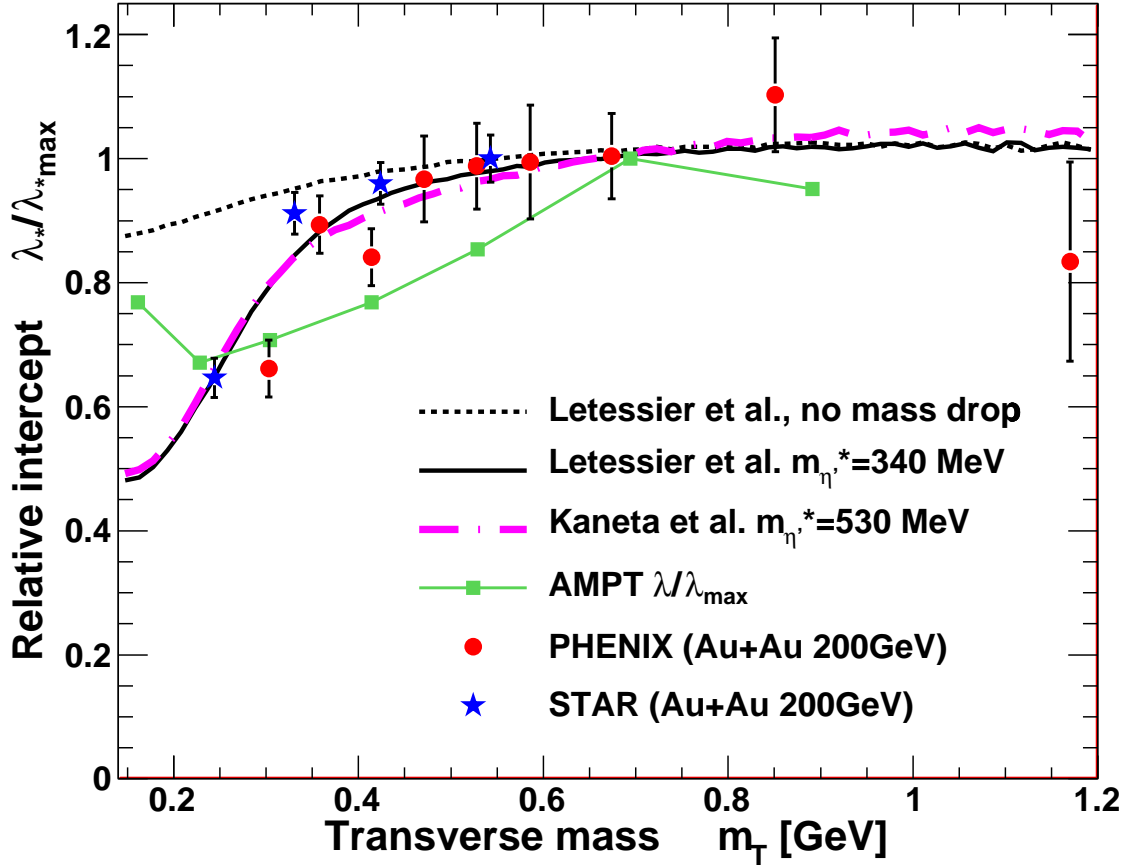}
\end{center}
\vspace{-0.5cm}
\caption{
Monte Carlo simulations of \lamfrac\ compared with PHENIX and STAR
data. The fits utilizing resonance multiplicities of refs.~\cite{rafelski,kaneta} (solid and dashed lines, respectively)
are compared to the no-mass-drop scenario of ref.~\cite{rafelski} (dotted line).
AMPT 2.11 $\lambda(\mT)/\lambda^\mathrm{max}$ simulation with string melting~\cite{Lin:2004en}  
report a non-thermal scenario without \etap mass modification, and its comparison to the data yields $\chi^2/ndf=102/13$, corresponding to $\rm{CL}=6.8 \times 10^{-16}$.
}
\label{fig:fit}
\end{figure}

Considering the \lamfrac distribution, shown in Fig.~\ref{fig:fit}, a dip is observable in measured RHIC $\sqsn=200\ \GeV$ central Au+Au collision data at low $\mT$ values.
(The \lambdas values used in this analysis and their total errors are discussed in details in ref.~\cite{Vertesi:2009wf}. Here $\lambdas^\mathrm{max}$ is the $\lambdas(\mT)$ value taken at $\mT=0.7\ \GeV$, with the exception of the STAR data, where the data point at the highest $\mT=0.55\ \GeV$ is considered. Note that the \mT dependency of the  $\lambdas(\mT)$ measurements in the 0.5-0.7 GeV region is very weak.)

\begin{figure}[hbtp]
\begin{center}
\includegraphics[width=.9\columnwidth]{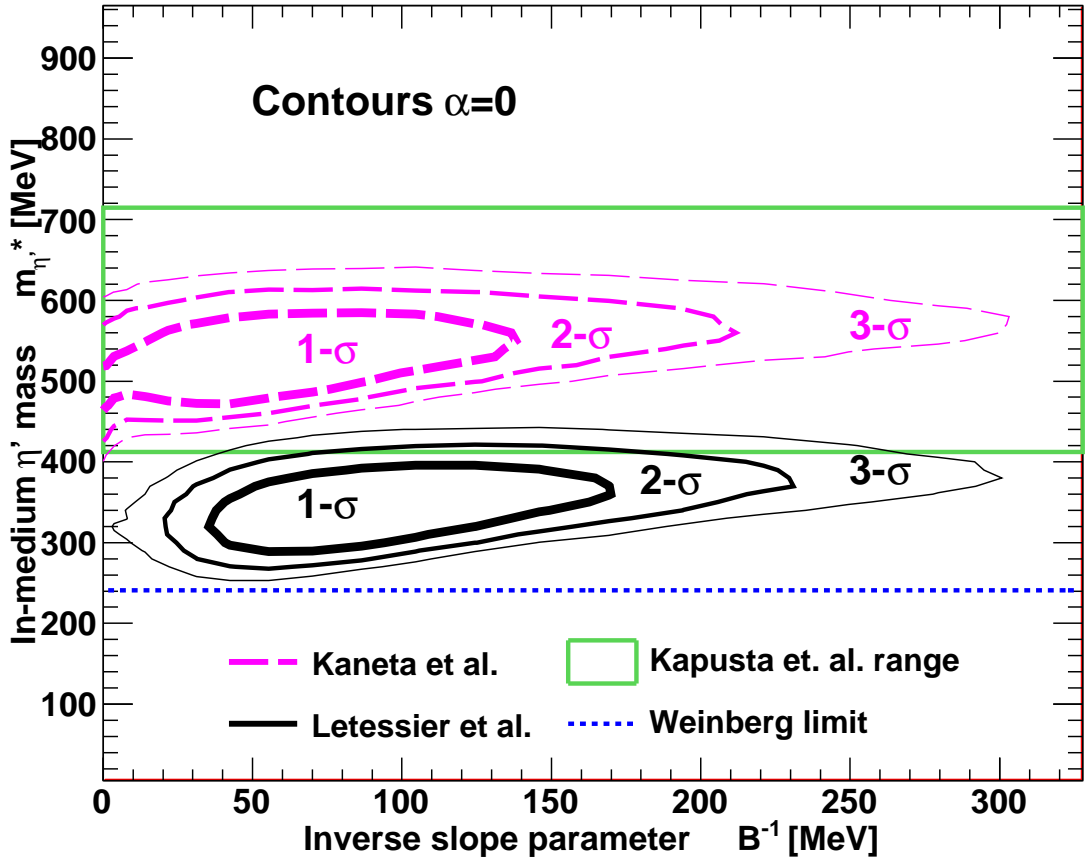}
\end{center}
\vspace{-0.5cm}
\caption{
Standard deviation contours on the (\Binv{}, \meps) plain, obtained from 
\lamfrac\ of Monte Carlo simulations based on particle multiplicities of refs.~\cite{kaneta,rafelski}, each fitted simultaneously to the PHENIX and STAR combined dataset.
The region between the horizontal solid lines indicates the range predicted in ref.~\cite{kapusta}.
The dotted horizontal line stands for Weinberg's lower limit~\cite{Weinberg:1975ui}.
}
\label{fig:contour}
\end{figure}

\begin{figure}[htbp]
\begin{center}
\includegraphics[width=\columnwidth]{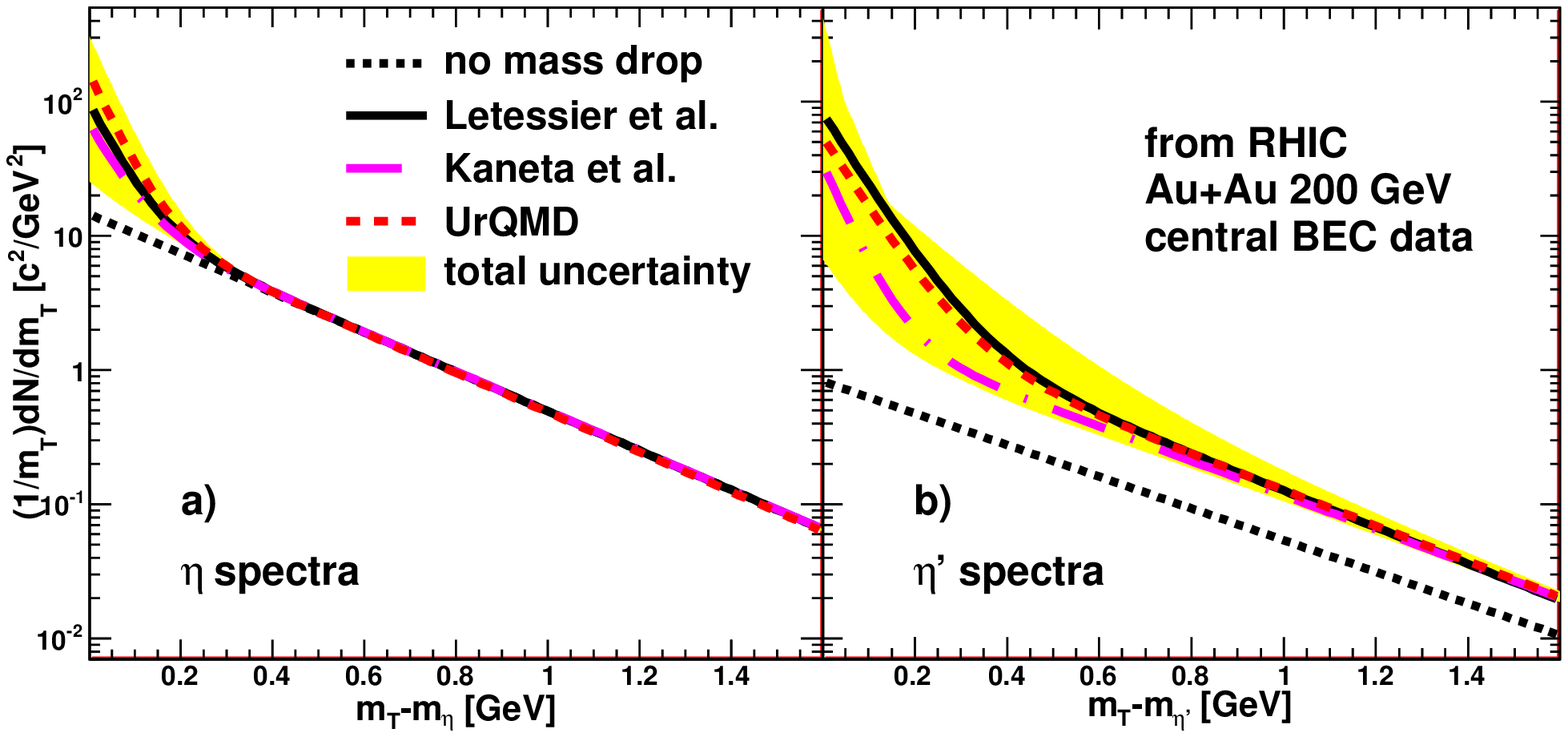}
\end{center}
\vspace{-0.5cm}
\caption{
{\it a)} Reconstructed $\mT$ spectrum of the $\eta$ and {\it b)} the \etap mesons.
The dotted line indicates the scenario without an in-medium \etap mass reduction, while the dashed, dot-dashed and solid lines show the enhancement required to describe the dip in the low \mT region of \lambdas corresponding to the resonance multiplicities of refs.~\cite{kaneta,urqmd,rafelski} respectively. 
The shaded regions indicate the total uncertainty of the reconstruction. Normalization was carried out with respect to the $\eta$ multiplicity of the model described in ref.~\cite{kaneta}.
}
\label{fig:spectrum}
\label{fig:etaspectrum}
\end{figure}

We have investigated a broad class of models of resonance production, 
including three different models that produce resonances without assuming local thermalization: 
AMPT, FRITIOF and UrQMD. 
AMPT, known to be fairly successful in describing the HBT radii without assuming any in-medium mass modification~\cite{Lin:2004en}, is not able to describe the dataset shown in Fig.~\ref{fig:fit}. 
The trend seen in AMPT can probably be attributed to a lower effective $\langle u_T\rangle$ of the high mass halo resonances~\cite{vance}.
(Note that $\lambda$, the Gaussian approximation of the intercept parameter reported by AMPT, is distinguished here from the extrapolated intercept \lambdas, which also includes the uncertainty from the fitting form~\cite{Vertesi:2009wf}.)
The FRITIOF~\cite{fritiof} Monte Carlo model, based on superposition of nucleon-nucleon collisions and the Lund string fragmentation model, cannot describe the behavior seen in \lamfrac even when an arbitrary \etap mass modification is considered. On the other hand, hadronic cascade based UrQMD~\cite{urqmd}, as well as the quark coalescence model ALCOR~\cite{alcor} and the thermal resonance production models of refs.~\cite{kaneta,stachel,rafelski}, provide a successful fit in a certain range of the in-medium \etap masses.
The main difference between the thermal models that we utilized is in those resonance multiplicities that are not yet measured well: ref.~\cite{kaneta} predicts a factor of 1.6 more $\eta$-s and a factor of 3 more $\etap$-s than the models of ref.~\cite{rafelski,stachel}. The relevant resonance fractions of these models are detailed in Table V of ref.~\cite{Vertesi:2009wf}.
Resonance decays, including decay chains, were simulated with JETSET 7.4~\cite{Sjostrand:1995iq}.

Based on extensive Monte-Carlo simulations, $\chi^2$ of the fits to the data of Fig.~\ref{fig:fit} was computed as a function of $m_{\etap}^*$ and $B^{-1}$ for each resonance model and each fixed value of model parameters of $\alpha$, $\Tcond$, $\Tfo$ and $\uT$. 
The best values for the in-medium mass of $\eta^\prime$ mesons are in, or slightly below, the range 
$\sqrt{\frac{1}{3}(2 m_{\rm K}^2 + m_\pi^2)}\le\meps\le \sqrt{2 m_{\rm K}^2 - m_\pi^2}$
predicted in  ref.~\cite{kapusta}, while all are above the lower limit of 
$\meps\ge\sqrt{3}m_\pi$ 
given by ref.~\cite{Weinberg:1975ui}.
The \lamfrac simulations for the best fits of two characteristic models are compared to the no-mass-drop scenario on Fig.~\ref{fig:fit}, while the 1, 2 and 3-$\sigma$ parameter boundaries are indicated in Fig.~\ref{fig:contour}. 
Those models that describe both PHENIX and STAR \lamfrac data in a statistically acceptable manner with the assumption of a sufficiently large in-medium $\eta^\prime$ mass reduction are all used for the estimation of systematics. The key parameters of the best fits are listed in Table~\ref{tab:modelsum}.
%
\begin{table}[htbp]
\begin{tabular}{c c c c c c}
\hline
Resonance & \meps & $\chi^2$ (CL \%) & \multirow{2}{*}{$f_\etap$} & \multirow{2}{*}{$f_\eta$} & 5-$\sigma$ limit \\
  model & (MeV) & $ndf$=11 &  &  & $\meps$ (MeV) \\[0.5ex]
\hline\hline
ALCOR~\cite{alcor}
& $490{+60\atop-50}$ & 20.2 (4.29) & 43.4 & 5.25 & $\le$ 700 \\
Kaneta~\cite{kaneta}
& $530{+50\atop-50}$ & 22.8 (4.12) & 25.6 & 3.48 & $\le$ 730 \\
Letessier~\cite{rafelski}
& $340{+50\atop-60}$ & 18.9 (6.35) & 67.6 & 4.75 & $\le$ 570 \\
Stachel~\cite{stachel}
& $340{+50\atop-60}$ & 18.8 (6.38) & 67.6 & 4.97 & $\le$ 570 \\
UrQMD~\cite{urqmd}
& $400{+50\atop-40}$ & 19.0 (6.14) & 45.0 & 7.49 & $\le$ 660 \\ [1ex]
\hline
\end{tabular}
\caption{\label{tab:modelsum}
Most probable fits of \meps for different resonance multiplicity models with the corresponding integrated enhancement factors $f_\etap$ and $f_\eta$ of the \etap and $\eta$ spectra respectively.
The errors on the \meps values represent the 1-$\sigma$ boundaries of the fits.
The 5-$\sigma$ limits of maximum in-medium masses including systematics are also shown. 
The fitted inverse slope parameters are $42\le \Binv \le 86$ for each model.
}
\end{table}

{\it Results: }
We have used different input models and setups to map the parameter space for a twofold goal. {\it i)} We excluded certain regions where a statistically acceptable fit to the data is not achievable, thus we can give a lower limit on the \etap mass modification. 
At the 99.9 \% confidence level, corresponding to a more than 5-$\sigma$ effect, at least 200 MeV in-medium decrease of the mass of the $\eta^\prime(958)$ meson was needed to describe both STAR 0-5 \% central and PHENIX 0-30\% central Au+Au  data on $\lamfrac$ in $\sqrt{s_{NN}} = 200$ GeV Au+Au collisions at RHIC, in the considered model class.
{\it ii)} We have determined the best values and errors of the fitted \meps and \Binv parameters. The best simultaneous description of PHENIX~\cite{phnxpub} and STAR~\cite{starpub} 
relative intercept parameter data is achieved with an $\etap$ mass that is dramatically reduced 
in the medium created in central Au+Au collisions at RHIC from its vacuum value of 958 MeV to
$340{+50\atop -60}{+280\atop -140}\pm{45}$ MeV.
The first error here is the statistical one determined by the 1-$\sigma$ boundaries of the fit. The second error is from the choice of the resonance model and the parameters ($\alpha$, $\Tcond$, $\Tfo$ and $\uT$) of the simulation. The third error is the systematics resulting from slightly different PHENIX and STAR centrality ranges, particle identification and acceptance cuts. These effects have been estimated with Monte-Carlo simulations, detailed in ref.~\cite{Vertesi:2009wf}, not to exceed 9.8\%, 7\% and 3\% respectively.
The main source of systematic errors is the choice of the resonance models. This is due to 
the unknown initial $\etap$ multiplicity, hence models like ref.~\cite{kaneta} with larger initial \etap abundances require smaller in-medium \etap mass modification, as compared to the models of ref.~\cite{stachel,rafelski}. 

In addition to the characterization of the in-medium $\etap$ mass modification, the transverse momentum spectra of the $\eta$ and $\etap$ mesons have also been determined. These spectra may serve as controls and provide motivation for future measurements as well as an input for theoretical calculations that may go well beyond the scope of the present manuscript.
The low transverse momentum enhancement of the \etap and $\eta$ spectra corresponding to the best fits is shown in Fig.~\ref{fig:spectrum}. Let us note that the enhancement of the $\eta$ production affects the $\pT\le 1\ \GeV$ region only. Our results do not modify the agreement of resonance models with the measured $\eta$ spectrum in the $\pT\ge 2\ \GeV$ region~\cite{Adler:2006bv}.

{\it Discussion:}
Detailed analysis of the STAR and PHENIX \lamfrac dataset recorded at 7.7, 9.2, 11.5, 39 and 62.4 GeV during 2010 has just been started~\cite{Abelev:2009bw}, marking the beginning of the RHIC energy scan program.

At present, detailed data are available from the NA44 collaboration at $\sqsn=19.4\ \GeV$~\cite{Beker:1994qv} as well as from the STAR collaboration at $\sqsn=62.4$ and 200 \GeV Cu+Cu and Au+Au collisions, the latter at different centrality classes within the 0\%--80\% range~\cite{Abelev:2009tp}. 
The NA44 data 
does not feature an \etap mass drop effect. A positive sign of the \etap mass modification is apparent in each case of the STAR datasets, indicating that the mass modification effect is nearly at maximum in $\sqsn=200\ \GeV$ Au+Au collisions and reduces with decreasing centrality, colliding energy and system size. We have estimated the magnitude of the system size and energy dependence between 62.4 GeV Cu+Cu and 200 GeV Au+Au collisions to be not larger than 15\%, which is substantially less than the dominant systematic error coming from the choice of the resonance model. 

Yet to be discovered {\it high mass resonances} might also lead to an enhancement of the soft pion production. Indeed, such an alternative scenario can successfully explain the energy dependence of the ${\rm K}^+/\pi^+$ and ${\rm K}^-/\pi^-$ ratios in relativistic heavy ion collisions~\cite{Chatterjee:2009km}. 
We have numerically tested the stability of our results for the enhancement of the pion halo coming from large mass long-lived resonances by switching on and off the contribution of the $\phi(1020)$ meson to estimate an upper limit of 2\% on the possible effect of exotic high mass resonances.

The {\it dilepton spectrum} has been measured recently in minimum bias Au+Au collisions at \sqsn = 200 GeV,
and a large enhancement was observed in the low invariant mass region $m_{\rm ee}<1\ \GeV$~\cite{Adare:2009qk}.
Low transverse mass enhancement of the \etap and $ \eta$ production results in dilepton enhancement just in this kinematic range~\cite{kapusta}. 
Estimations using the enhancement factors in Table~\ref{tab:modelsum} indicate that the observed in-medium \etap mass drop is indeed a promising candidate to explain this dilepton excess.

PHENIX recently reported a two-component transverse momentum spectrum in dilepton channel {\it direct photon measurements}~\cite{Adare:2009qk}, which provides an additional testing possibility to constrain the two component structure of the \etap and $\eta$ spectra reconstructed here.

{\it In summary}, we report on a statistically significant, indirect observation of an in-medium mass modification of the $\eta^\prime$ mesons in $\sqsn=200\ \GeV$ Au+Au collisions at RHIC.
A similar search for in-medium $\etap$ mass modification provided negative result in S+Pb reactions at
CERN SPS energies~\cite{vance}.
Further, detailed studies of the excitation function, the centrality and system size dependence of the $\lamfrac$ could provide important additional details about the onset and saturation of the partial \UA symmetry restoration in hot and dense hadronic matter. 
Studies of the low-mass dilepton spectrum and measurements of other decay channels 
of the $\etap$ meson may shed more light on the reported magnitude of the low $\pT$ $\etap$ enhancement and the related \UA symmetry restoration in high energy heavy ion collisions.
%

\begin{acknowledgments} 
We thank R.~J.~Glauber and Gy.~Wolf for inspiring discussions. T.~Cs. is grateful to R.~J.~Glauber for his kind hospitality at the Harvard University. Our research was supported by Hungarian OTKA grant NK 73143, and by the Hungarian American Enterprise Scholarship Fund (HAESF).
\end{acknowledgments}

\end{document}